\documentclass[reprint,amsmath,amssymb]{revtex4-2}
\usepackage{graphicx}
\usepackage{dcolumn}
\usepackage{bm}
\usepackage[utf8]{inputenc}
\usepackage{footmisc}
\graphicspath{ {./images/} }
\usepackage[mathlines]{lineno}

\begin{document}

\preprint{APS/123-QED}

\title{Tuning magnetocaloric effect by optimizing thickness induced 3D strain state}

\author{Samir Kumar Giri$^{1!}$}

\author{Wasim Akram$^{2!}$}
\author{Manisha Bansal$^{2}$}
\author{Tuhin Maity$^{2}$}
 \email{tuhin@iisertvm.ac.in\\
 \\
! Equal Contribution first authors}
 \affiliation{$^{1}$Kharagpur College, Kharagpur, Paschim Medinipur, West Bengal 721305, India\\
$^2$School of Physics, Indian Institute of Science Education and Research Thiruvananthapuram, Thiruvananthapuram, Kerala 695551, India}

\date{\today}

\begin{abstract}
The effect of 3-dimensional strain state on the magnetocaloric properties of epitaxial La$_{0.8}$Ca$_{0.2}$MnO$_3$ (LCMO) thin films grown on two types of substrates, 
SrTiO$_3$ (001) (STO)  and LaAlO$_3$ (001) (LAO)  has been studied as a function of film thickness within the range of 25 – 300 nm. The STO substrate imposes an in-plane tensile biaxial strain while LAO substrate imposes an in-plane compressive biaxial strain. The in-plane biaxial strain on LCMO by STO substrate gets relaxed more rapidly than that by LAO substrate but both LCMO/STO and LCMO/LAO show a maximum entropy change ($\Delta{S}_M$) of $\sim$ 12.1 JKg$^{-1}$K$^{-1}$ and $\sim$ 3.2 JKg$^{-1}$K$^{-1}$, respectively at a critical thickness of 75 nm (at 6 T applied magnetic field). LCMO/LAO is found to exhibit a wider transition temperature region with full width at half maxima (FWHM) $\sim$ 40 K of the $\frac{dM}{dT}$ vs $T$ curve compared to LCMO/STO with FWHM $\sim$ 33 K of that curve. This broadening of the transition region  indicates that the Table like magnetocaloric effect (MCE) is attainable by changing the strain type. The maximum Relative Cooling Power, $\sim$ 361 JKg$^{-1}$ of LCMO/STO and $\sim$ 339 JKg$^{-1}$ of LCMO/LAO is also observed at the thickness $\sim$ 75 nm. The Curie temperature varies with the thickness exploring the variation of ferromagnetic interaction strength due to strain relaxation. The film thickness and substrate induced lattice strain are proved to be the significant parameters for controlling MCE. The highest MCE response at a particular thickness shows the possibility of tuning MCE in other devices by optimizing thickness.

\end{abstract}

\maketitle
\section{Introduction}

Depletion of energy resources is pushing the scientific community to materialize the energy-efficient and environmental friendly technologies. To sustain the nature and the running demands alongside, low power refrigeration or cooling at both macro and micro scale has been proved to be one of the future energy saving innovation\cite{gschneidner1999recent,wang2018outstanding,LI2020153810,gottschall2019making}. Existing cooling techniques include absorption and adsorption refrigerators, thermoelectric cooling, thermoacoustic refrigerators, ejector refrigeration systems, magnetocaloric (MC) refrigeration, etc\cite{kitanovski2020energy,moya2014caloric}. Among these, magnetocaloric effect (MCE) is the most promising candidate due to its efficiency and environmental friendly approach. Use of thermal effects induced by the application of magnetic field to produce efficient cooling, popularly known as MCE has gained recognition among the researchers\cite{moya2015too,moya2020caloric,FRANCO2018112}. Room temperature cheap and efficient MC devices lead to potent energy safeguarding household and industrial applications outperforming conventional cooling or refrigeration techniques\cite{tegus2002transition}. However, only few materials can be used for MC applications due to the complexity involved in studying the coupled magnetic and structural parameters of the system\cite{pecharsky1997giant,pecharsky1997tunable,wada2001giant}. Materials should have high MC constant, wide range of operating temperature, low hysteresis loss, low specific heat, high thermal conductivity, etc., to produce large MCE response which is challenging as of significant technological importance. Recent material researches show possible improvement in the MCE efficiency concomitant with challenges like fabrication complexity, less resistance to corrosion, poor conductivity, etc\cite{wang2018outstanding,lyubina2012novel,lyubina2010novel}. Different alloys (e.g., Iron based alloy Gd$_5$Ge$_{1.9}$Si$_2$Fe$_{0.1}$)\cite{provenzano2004reduction} and composites (e.g., La(Fe, Si)$_{13}$Hy/In)\cite{wang2018outstanding} showed potential to overcome such deficiencies, henceforth improving the performance of MCE based magnetic refrigeration or cooling.

Moreover, most research activities on MCE are limited to bulk materials since the studies on thin films are more complex due to the intricacies involved in their fabrication and study. But the industries especially for micro and nanoscale devices are in a huge demand of thin film based MCE because of their high efficiency, less space consumption, flexibility, larger specific surface area for higher exchange of heat, etc\cite{khovaylo2014magnetocaloric,belo2019magnetocaloric}. Interestingly, epitaxial thin films show several fascinating phenomena including colossal magnetoresistance (CMR) due to the additional strain constraints\cite{PhysRevLett.76.1356,jin1994colossal} which are of particular importance to determine their physical properties\cite{choi2020nanoengineering,zarifi2016effects,baena2011effect,mukherjee2013theory}. The giant MCE has been observed in the CMR based epitaxial manganite systems like La$_{0.67}$Sr$_{0.33}$MnO$_3$ (LSMO)/BaTiO$_3$ (BTO) ($\Delta{S_M}/\mu_0\Delta{H}$ = 1.95 JKg$^{-1}$K$^{-1}$T$^{-1}$)\cite{giri2019strain}, La$_{0.7}$Ca$_{0.3}$MnO$_3$/BTO ($\Delta{S_M}/\mu_0\Delta{H}$ = -0.7 JKg$^{-1}$K$^{-1}$T$^{-1}$ for intrinsic and -9.0 JKg$^{-1}$K$^{-1}$T$^{-1}$ for extrinsic)\cite{moya2013giant}, where the substrate induced strain influences the temperature dependent magnetic properties of the CMR materials resulting in adiabatic thermal process\cite{phan2007review,amaral2005magnetocaloric,wang2001magnetocaloric}. This substrate induced strain may be compressive (tensile) depending upon larger (smaller) lattice parameter of the epitaxial film than that of the substrate\cite{antonakos2007strain,zarifi2016effects,aydogdu2008novel} which results in decrease (increase) of the Mn-O bond length as well as altering the Mn-O-Mn bond angle\cite{zarifi2016effects}, enhancing MCE. Generally, under a tensile strain, ferromagnetism due to double exchange (D-E) interaction gets weakened because of the reduction in $e_g$ electron transfer integral, $t_\alpha$ = $t_\alpha^{\star}$sin($\phi$/2) where $t_\alpha^{\star}$ is the bare transfer integral and $\phi$ is the Mn–O–Mn bond angle\cite{zhang2001anomalous} and decreases the Curie temperature ($T_C$)\cite{zhang2001anomalous}. The enhancement of $T_C$ under compressive strain\cite{razavi2000epitaxial} can also be explained by the same mechanism. So, it’s expected that the strain relaxation by increasing film thickness ($t$) should bring about the $T_C$ to higher values always for the tensile strain and to lower values for the compressive strain. On the contrary, several reports\cite{gong1996perovskite,prellier2000spectacular,kanki2001anomalous} were able to convey that the tensile strain also drives up $T_C$ while the compressive strain reduces it\cite{zhang2001strain} and relaxation of these strains would do the opposite as expected. 

In this work, we investigate the evolution of strain states on MCE of the epitaxial La$_{0.8}$Ca$_{0.2}$MnO$_3$ (LCMO) thin films grown on STO and LAO substrates which give rise to tensile and compressive strain, respectively and show the improvement of performance by large tuneable magnetic entropy change as well as the variation of transition temperatures. It is observed that the transition temperatures do not always decrease or increase under strain but varies with $t$ witnessing a maximum value at 75 nm for both the tensile and compressive strain. For cooling applications, distributed magnetic ordering temperature is highly desiderated as it brings on to “Table like MCE”\cite{chaturvedi2011table,singh2005anomalous} which is the basis of ideal Ericsson-cycle based magnetic refrigeration\cite{chaturvedi2011table}. Here we report the broadening of the transition region for LCMO/LAO than that of the LCMO/STO suggesting the possibility of acquiring “Table like MCE”, a goal towards the ideal Ericsson-cycle. To compare the viability of these LCMOs grown on STO and LAO as an efficient refrigerant, the relative cooling power (RCP) is calculated and plotted against $t$ and magnetic field ($H$). The bulk La$_{0.67}$Ca$_{0.33}$MnO$_3$ shows a magnetic entropy change, $\Delta{S_M}$ = 0.9 JKg$^{-1}$K$^{-1}$T$^{-1}$\cite{zhang1996magnetocaloric} which underscores the need for efficient cooling application. Here, we show $\Delta{S_M}$ = 2 JKg$^{-1}$K$^{-1}$T$^{-1}$ in epitaxial  La$_{0.8}$Ca$_{0.2}$MnO$_3$ by optimizing $t$ and substrate induced lattice strain.
\section{Experimental method}
The epitaxial LCMO thin films were grown by pulsed laser deposition (PLD) technique from a stoichiometric target at 750$^o$C with an oxygen pressure of 300 mTorr and an estimated laser energy density $\sim$ 3.5 Jcm$^{-2}$ with a repetition rate of 5 Hz. Two different types of single crystal substrates i.e., LAO (001) (a$^P$ $\sim$ 3.79Å) and STO (001) (a$^P$ $\sim$ 3.905Å) were used to attain two different types of lattice mismatch for the growth of LCMO thin films. The thicknesses of the films were varied from 25 nm to 300 nm in order to correlate the strain relaxation process with MCE. The composition of the film with target was confirmed from wavelength dispersive x-ray spectroscopy within experimental error.
Structural characterization was carried out using four-circle x-ray diffractometer. Figs. 1(a) and 1(b) show XRD scans of 25 nm LCMO films grown on STO and LAO substrates, respectively. The  out-of-plane and in-plane lattice parameters were determined by normal $\theta-2\theta$ and grazing incidence diffraction (GID) scans at room temperature, respectively. The variation of in-plane lattice parameters w.r.t. $t$ for the two different substrates is shown in Figs. 1(c) and 1(d). The bulk value of LCMO lattice parameter (3.881 Å) is obtained from the target pellet. 
Temperature-dependent magnetizations of all LCMO thin films were measured, in order to explore the effect of 3D strain states on magnetic properties. Magnetization measurements were carried out using SQUID magnetometer in the temperature range 5-300 K and field range of ±6 T. The temperature dependent zero field cooled (ZFC) Magnetization ($M$) vs Temperature ($T$) of all the films are taken at $H$ = 5 mT. For slow magnetic field scan, 0.01 T per minute ramp rate was used in SQUID to consider all the magnetization process as quasi equilibrium and isothermal.

\begin{figure}
\centering
\includegraphics[width=9cm]{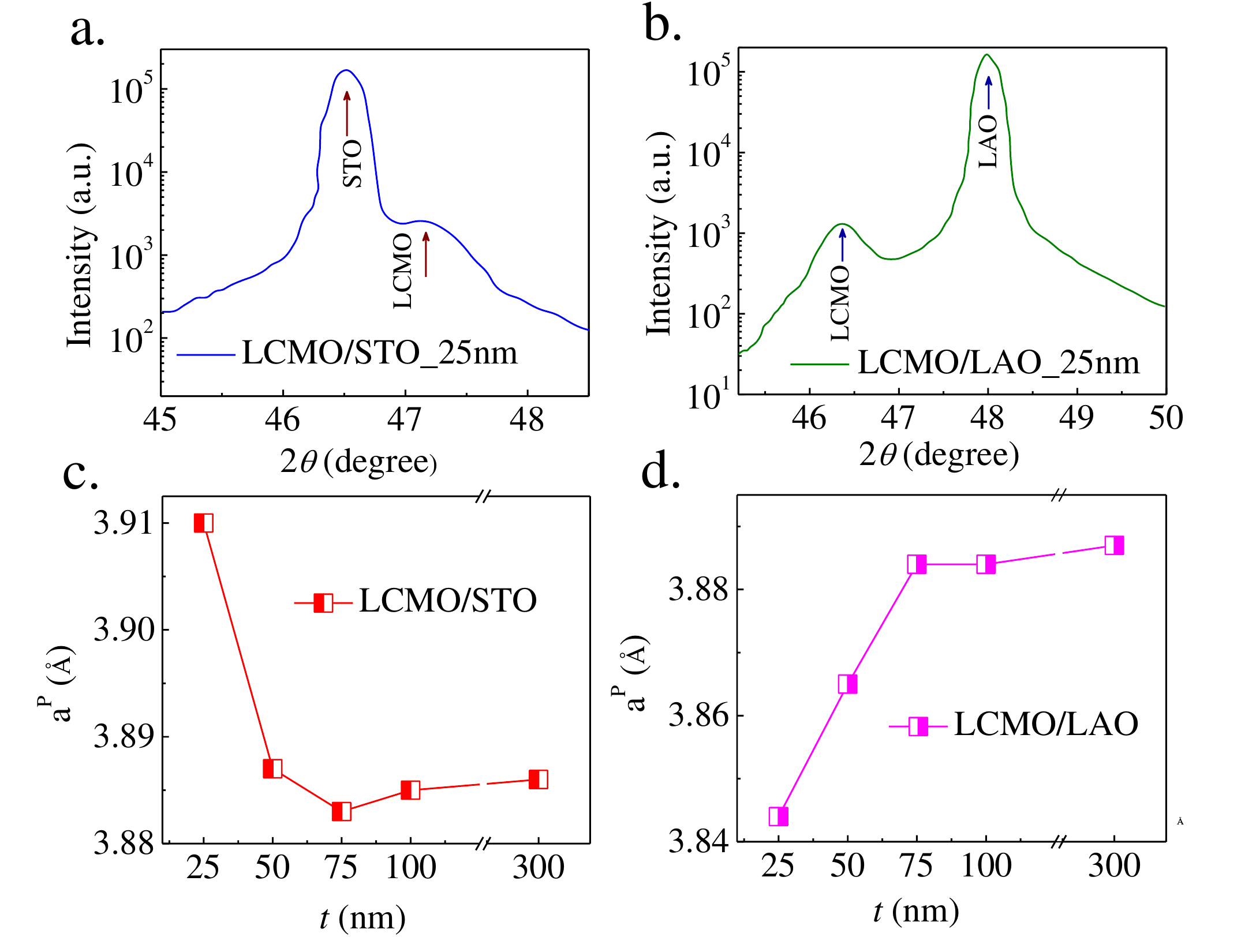}
\caption{(a) and (b) Grazing incidence XRD scans of 25 nm LCMO films grown on STO and LAO substrates, respectively. (c) and (d) show the change of lattice parameters as a function of $t$ of LCMO/STO and LCMO/LAO, respectively}
\end{figure}
\section{Results and Discussion}
It’s evident from the XRD scans that the epitaxially grown LCMO thin films are subjected to two distinct strain states on LAO and STO substrates. In Figs. 1(c) and 1(d), the in-plane lattice parameters (a$^P$) of the LCMO thin films obtained from grazing incidence x-ray diffraction (GID) at room temperature are found to be deviated from that of the substrates to the bulk value (3.881Å) with the increase in $t$. Rao \emph{et al.}\cite{rao1999effects} have previously discussed the variation of 3D strain state w.r.t. $t$ in detail. LAO substrate (a$^P$ $\sim$ 3.79Å) imposes an in-plane compressive strain with lattice mismatch -2.34 \% while STO substrate (a$^P$ $\sim$ 3.905Å) imposes a corresponding tensile strain with a lattice mismatch +0.62 \% on the films.

The temperature dependence of the in-plane magnetization of LCMO/STO and LCMO/LAO are shown in Figs. 2(a) and 2(b), respectively. The transition temperatures are obtained from the extrema of $\frac{dM}{dT}$ vs $T$ curves (insets of Figs. 2(a) and 2(b)). LCMO/STO shows a cusp at $T_P$ in lower temperature regime as well as at Curie temperature $T_C$ in higher temperature regime. The full width at half maxima (FWHM) of $\frac{dM}{dT}$ vs $T$ curve for LCMO/LAO ($\sim$ 40 K) is larger than that of the LCMO/STO ($\sim$ 33 K) (Fig. 2(c)) elucidating the broadening of the transition region for the former compared to the latter. Along with the absence of any thermal hysteresis in $MT$ curve for LCMO/LAO, this suggests the probability of second order magnetic phase transition (SOMPT) while the thermal hysteresis in $MT$ curve for LCMO/STO (Fig. 2(d)) between field cooling and field heating elucidates the first order magnetic phase transition (FOMPT) at transition temperature\cite{thanh2018tuning}. 
\begin{figure}
    \centering
    \includegraphics[width=8.8cm]{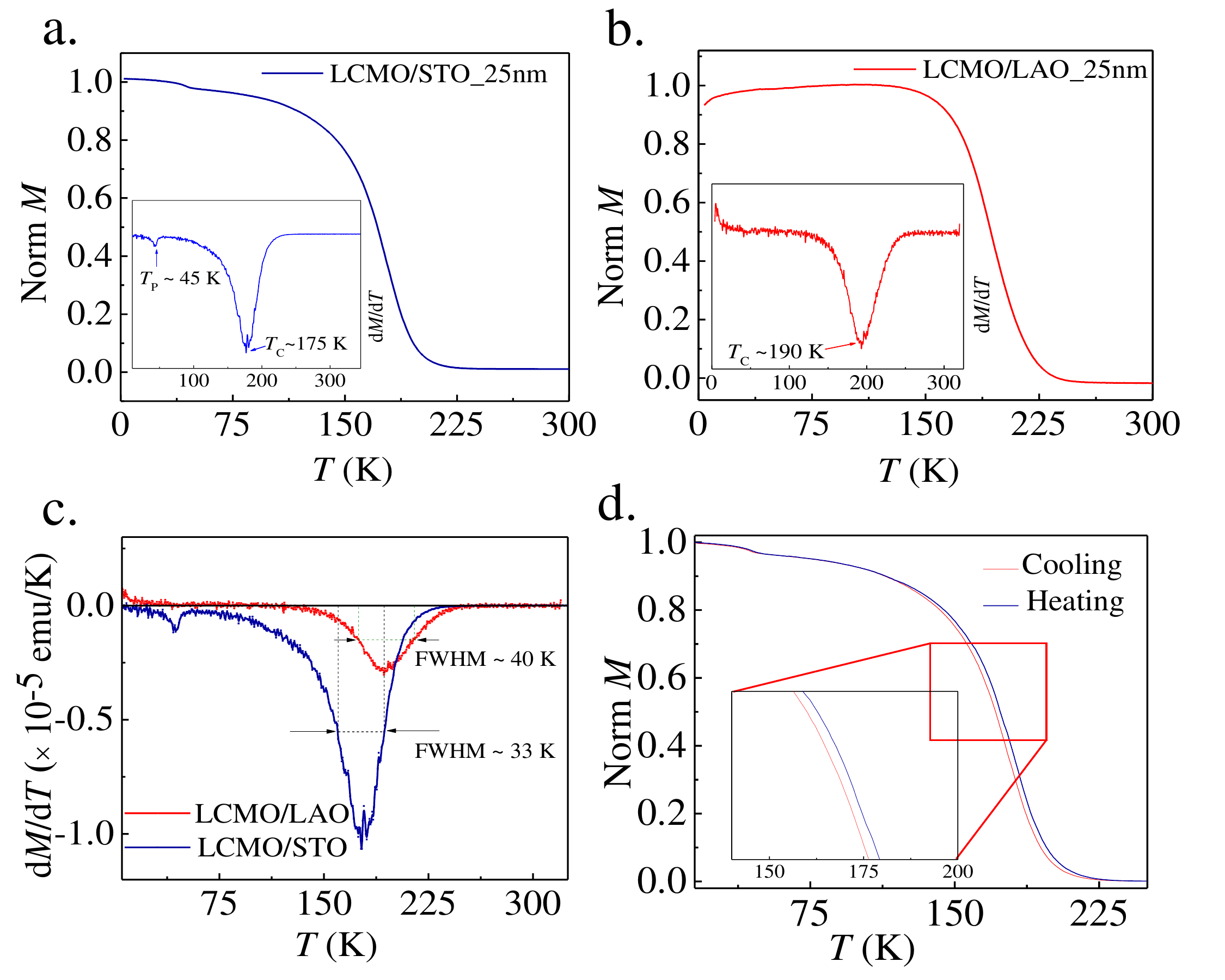}
    \caption{(a) and (b) $M$ vs $T$ curves of 25 nm LCMO thin film grown on STO and LAO substrates, respectively under 5 mT magnetic field. Inset figures show the $\frac{dM}{dT}$  vs $T$ curves and corresponding $T_C$. (c) Broadening of transition region for LCMO/STO compared to LCMO/LAO. (d)  Hysteresis between warming up and cooling down cycle of magnetization for LCMO/STO under 5 mT field. }
\end{figure}

For comparison, $T_C$ and $T_P$ of the epitaxial LCMO films are plotted as a function of $t$ in Fig. 3. LCMO/LAO shows higher $T_C$ compared to LCMO/STO of the same $t$ due to persistence of stronger ferromagnetism below transition temperature $T_C$ under compressive strain\cite{fu2000grain}. However, the transition temperatures $T_C$ and $T_P$ vary non-monotonously with the strain relaxation on increasing $t$ below 100 nm thickness. Both of them at first go on increasing for up to 75 nm thickness and then decrease. This can be explained as dominating behavior of ferromagnetic (FM) phase up to 75 nm thickness which has been discussed later. At 300 nm thickness both $T_C$ and $T_P$ are higher which is usual for thin film\cite{zhang2001thickness,almahmoud2010dependence}. The increase in $T_C$ for tensile strain relaxation up to $t$ $\sim$ 75 nm originates from the strengthening of D-E interaction due to Mn-O bond length reduction\cite{zhang2001anomalous}. However, the compressive strain induced ferromagnetism in LCMO system observed here is in contrast as $T_C$ was supposed to be decreased on strain relaxation due to weakening of D-E interaction. So, this anomaly needs to be understood. 

\begin{figure}
    \centering
    \includegraphics[width=7cm]{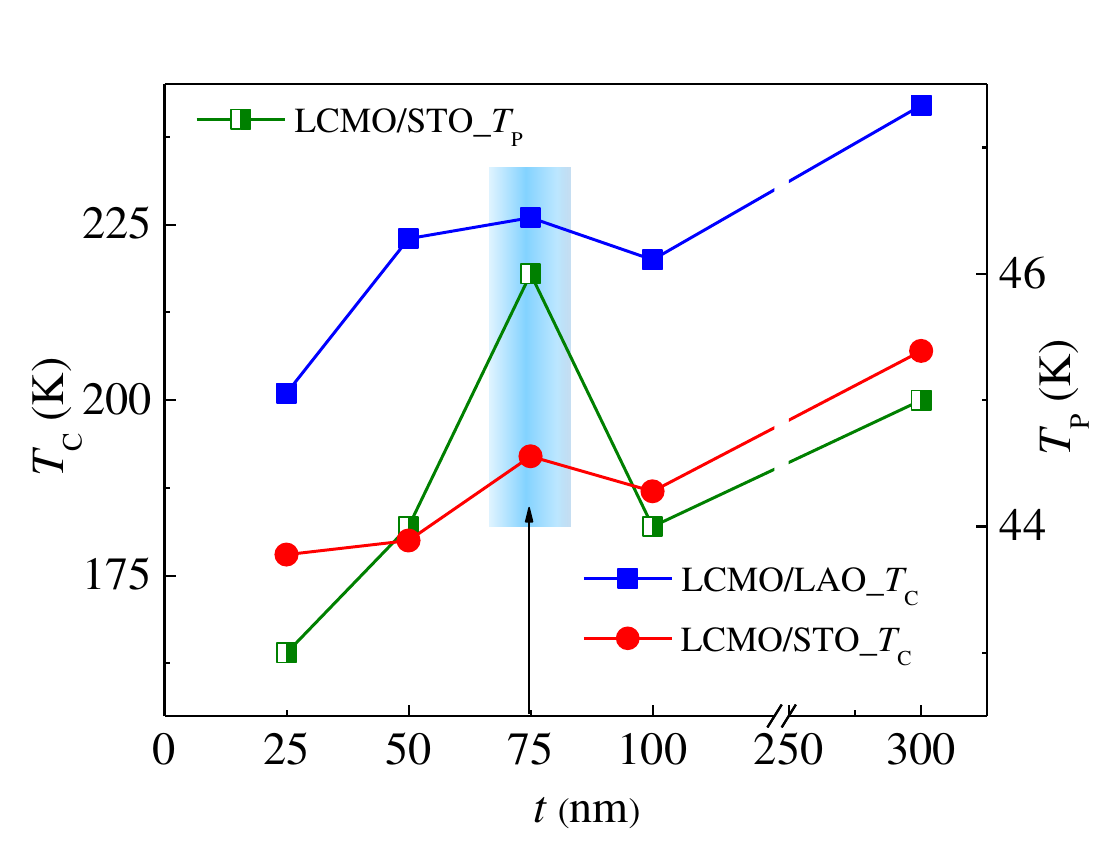}
    \caption{$T_C$ and $T_P$ vs $t$ plot for LCMO/LAO and LCMO/STO.}
\end{figure}

To understand this, two effects should be taken into consideration as the result of the in-plane Mn-O bond length elongation (contraction) – one is the decrease (increase) of $e_g$ electron transfer integral as mentioned before and the another is electron localization in $d_{x^2-y^2}$ -orbital\cite{zarifi2016effects} due to in-plane $d_{x^2-y^2}$ -orbital stabilization\cite{zhang2001anomalous} which can enhance the ferromagnetism and increase $T_C$. In particular, Mn$^{3+}$ ions in LaMnO$_3$ are already known to be arranged in an A type planar antiferromagnetic structure consisting of oppositely aligned FM \{001\} planes\cite{wang2015imaging}. Hence, in-plane Mn$^{3+}$ ions interact ferromagnetically with each other. When LaMnO$_3$ is doped with divalent Ca$^{2+}$ ions, they replace some of the trivalent La$^{3+}$ ions and form mixed valence state in Mn ions (Mn$^{3+}$ and Mn$^{4+}$)\cite{coey1999mixed}. The Mn$^{3+}$ and Mn$^{4+}$ ions interact with each other ferromagnetically through D-E interaction mediated by O-$2p$ orbital electrons\cite{goodenough1955theory}. The spacing between Mn$^{4+}$ and Mn$^{3+}$ gets increased under tensile strain giving rise to reduction\cite{zhang2001anomalous} in $e_g$ orbital electron transfer integral and weakening ferromagnetism due to D-E interaction and the reverse case happens for the compressive strain. But, at the same time, due to elongation of  the in-plane Mn-O bond length, the overlapping between the lobes of the in-plane Mn-$3d$ orbitals and O-$2p$ orbitals gets decreased\cite{cox1996electronic} giving rise to lower coulomb repulsion and favoring the in-plane  orbital electron localization\cite{zarifi2016effects} which in turn induces ferromagnetism. So, due to the competition between these two effects, the practical behavior depends on the dominant effect or possibly a mixed effect. For tensile strain, the reduction in D-E interaction dominates over   orbital electron localization whereas the latter one dominates over the first one for compressive strain and due to compressive strain relaxation, the in-plane   orbital electrons are getting localized to give rise to stronger ferromagnetism. Above 75 nm thickness, $T_C$ for each sample starts to fall from the maximum value indicating that the dominating behavior is\ not always the same throughout the strain relaxation process. Similarly $T_P$ achieves the highest value at 75 nm of thickness as well.

The viability of these epitaxial LCMOs as the core refrigeration system can be investigated by a comparative study of the MCE produced by these. The MCE is parameterized by the isothermal magnetic entropy change ($(\Delta{S}_M)_{iso}$) and/or adiabatic temperature change ($\Delta{T}_{ad}$) of a magnetic material under the variation of $H$\cite{belo2019magnetocaloric}. Most of the research reports publish $(\Delta{S}_M)_{iso}$ data instead of $\Delta{T}_{ad}$ for the former being easily measurable with standard magnetometers, but $\Delta{T}_{ad}$ measurement has some technical difficulty at the micro and nano scale due to uncertainty of adiabatic procedure and occurrence of the rapid thermal diffusion from the studied micro/nanostructure towards a heat/cold sink\cite{moya2014caloric,belo2019magnetocaloric}. The isothermal magnetic entropy change $\Delta{S}_M(H)$ of a magnetic material on application of magnetic field $H$ at temperature $T$ can be written from the Maxwell’s thermodynamic relation $(\frac{\partial S}{\partial H})_T$ = $(\frac{\partial M}{\partial T})_H$  as\cite{moya2013giant}

\begin{equation}
\begin{split}
\Delta{S}_M(T,H) & = {S}_M(T,H)-{S}_M(T,0)\\
 & = \int_{0}^{H} (\frac{\partial M}{\partial T})_H \,dH
\end{split}
\end{equation}
This expression ensures the maximum magnetic entropy change around the transition temperature due to switching in magnetic ordering and rapid change of magnetization. For these LCMO films, the values of $\Delta{S_M}$ are calculated from the isothermal magnetization curves (not shown here) obtained at different temperatures with appropriate interval of temperature $\Delta{T}$. Hence Eq. (1) can be approximated by 
\begin{equation}
\mid {\Delta{S_M}} \mid = \sum{\frac{(M_n-M_{n+1})}{(T_{n+1}-T_n)} \,{\Delta{H_n}}}
\end{equation}
where, $M_n$ and $M_{n+1}$ are the magnetization values measured in a field $H$ at temperatures $T_n$ and $T_{n+1}$, respectively\cite{amaral2010estimating,burrola2020tuning}. The values of temperature dependent $\Delta{S_M}$ are calculated by using Eq. (2) as shown in Fig. 4 for all LCMO/STO and LCMO/LAO. 
 \begin{figure}
\centering
\includegraphics[width=9cm]{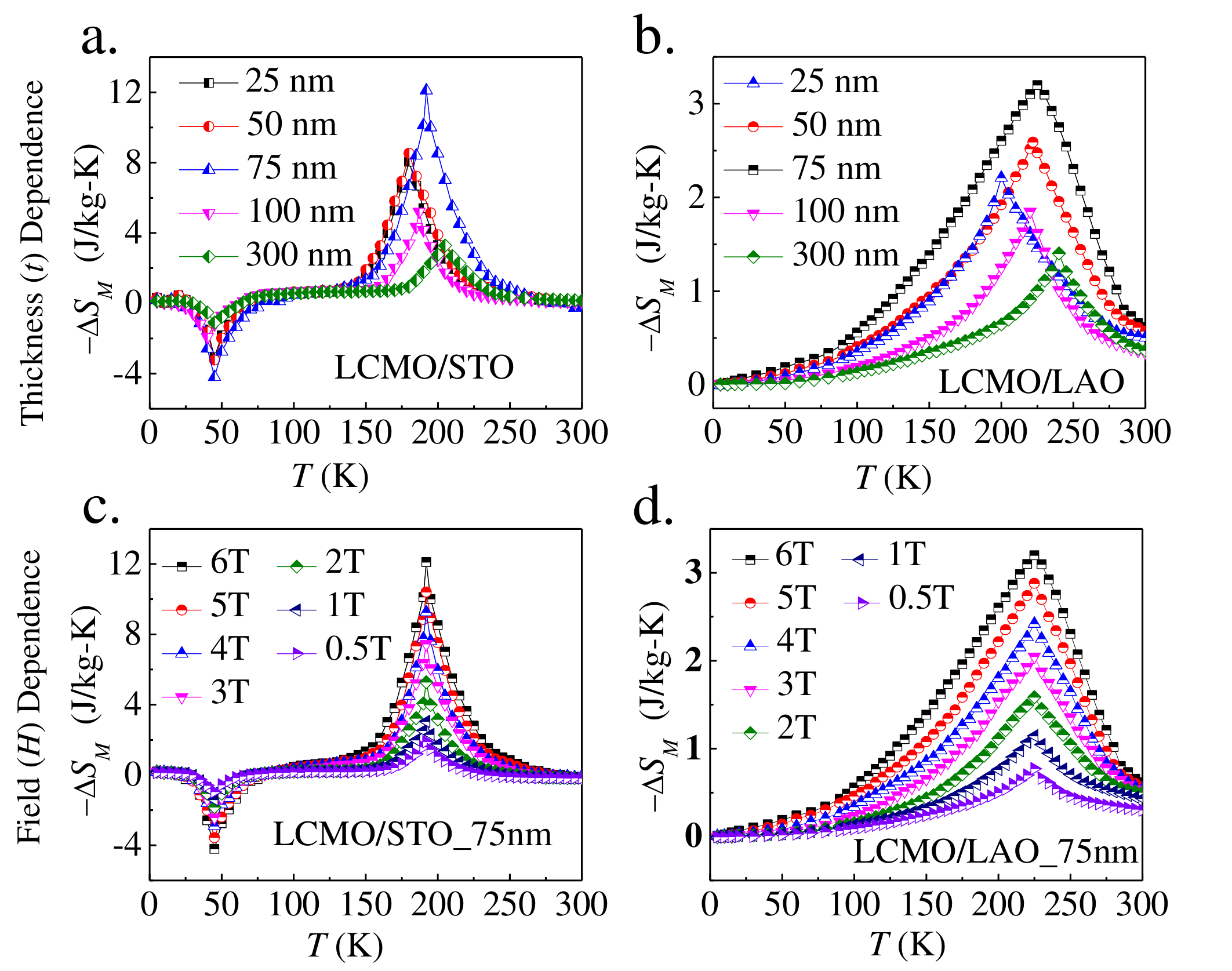}
\caption{$\Delta{S_M}$ vs $T$ curve for (a) LCMO/STO and (b) LCMO/LAO on varying $t$ from 25 nm to 300 nm. (c) and (d) are $\Delta{S_M}$ vs $T$ curve for LCMO/STO and LCMO/LAO, respectively on varying $H$ from 6 T to 0.5 T}
\end{figure}

Here, the maximum value of $\Delta{S_M}$ (Fig. 5(a)) is found to be $\sim$ 12.1 JKg$^{-1}$K$^{-1}$ for LCMO/STO of 75 nm thickness under a field change of 6 T. Interestingly, the value of $\Delta{S_M}$ changes due to variation of strain type on different substrates and strain amount of different $t$ (Fig. 5(a)). The values of $-\Delta{S_M}$ derived from the Maxwell equation show a negative peak (i.e., normal MCE) around $T_P$ (for LCMO/STO) and positive peak (i.e., inverse MCE) around $T_C$ (for LCMO/STO and LCMO/LAO both), respectively. Both the peaks increase with increasing $H$ (Figs. 4(c) and 4(d)). For LCMO/STO, the value of $-\Delta{S_M}$ (+ 12.1 JKg$^{-1}$K$^{-1}$) at $T_C$ is almost three times larger in magnitude than at $T_P$ ($-\Delta{S_M}$= - 4.2 JKg$^{-1}$K$^{-1}$) for $\mu_0H$ = 6 T and negative $\mid$$\Delta{S_M}$$\mid$ remains in narrow temperature regime around $T_C$ due to FOMPT while for LCMO/LAO, it gets broadened over a wide temperature range due to the suppression of FOMPT with the peak value of 3.2 JKg$^{-1}$K$^{-1}$. To make sure that the temperature for maximum entropy change ($T_{P2}$) and the transition temperature ($T_C$) follow the similar trend, $T_{P2}$ is plotted against $t$ (Fig. 5(b)). 

\begin{figure}
    \centering
    \includegraphics[width=8.8cm]{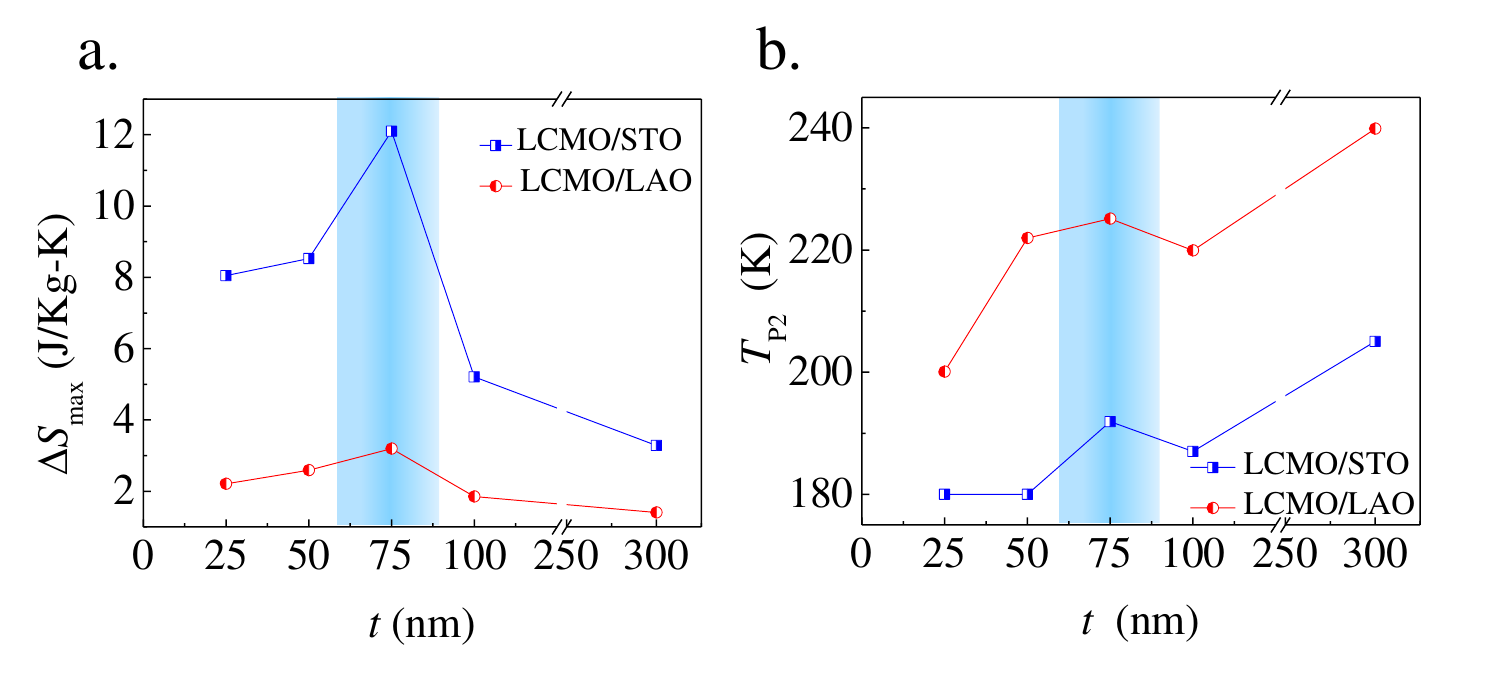}
    \caption{(a) Plot of $\Delta{S_{max}}$ against $t$ for LCMO/STO and LCMO/LAO (b) Temperature at $\Delta{S_{max}}$($T_{P2}$) vs $t$ plot }
\end{figure}
At 75 nm thickness both the $\Delta{S_{max}}$ and $T_{P2}$ show maxima and to investigate this anomalous behavior, we plotted the JT distortion ($\epsilon_{JT}$) and in-plane biaxial strain ($\epsilon_{XX}$ = $\epsilon_{YY}$) against $t$ (Figs. 6(a) and 6(b)). Both are decaying exponentially following the equation 
\begin{equation}
\epsilon = {\epsilon^\star}e^{-\frac{t}{t_0}} + \epsilon_0
\end{equation}
with the fitting parameters ${\epsilon^\star}$, $\epsilon_0$ and $t_0$ [for $\epsilon_{JT}$ curve (dark cyan colored) of LCMO/LAO (Fig. 6(a)) and LCMO/STO (Fig. 6(b)), ${\epsilon^\star}$ = 4.42 \% and -1.92 \%, $\epsilon_0$ = -0.3 \% and -0.42 \%, and $t_0$ = 30.9 nm and 25.5 nm,  respectively; for $\epsilon_{XX}$ curve (red colored) of LCMO/LAO (Fig. 6(a)) and LCMO/STO (Fig. 6(b), ${\epsilon^\star}$ = -2.25 \% and 3.71 \%, $\epsilon_0$ = 0 \% and 0 \%, and $t_0$ = 30 nm and 15.5 nm, respectively.] where $t_0$ is the thickness at which the strain drops to $1/e$ times of its maximum value and indicates how fast the strain gets relaxed with increasing $t$. The product ($\epsilon_{XX} \times \epsilon_{JT}$) which is the combined effect on $\Delta{S_{max}}$ by JT distortion and in - plane strain also follows the similar exponential decay curve
\begin{equation}
(\epsilon_{XX} \times \epsilon_{JT}) = (\epsilon_{XX} \times \epsilon_{JT})^{\star}e^{-\frac{t}{t_0}}
\end{equation}
with the fitting paramaters $(\epsilon_{XX} \times \epsilon_{JT})^{\star}$ = -0.052 \% and -0.075 \%, and $t_0$ = 20 nm and 11.5 nm for LCMO/LAO (Fig. 6(c)) and LCMO/STO (Fig. 6(d)), respectively. All the strain components  mentioned here and their product for LCMO/STO decay more rapidly than those of the LCMO/LAO, respectively.

For the given range, the $t$ dependence of $\Delta{S_{max}}$ can be well fitted with Lorentz function,
\begin{equation}
\Delta{S_{max}} = (\Delta{S_{max}})_0 + \frac{a}{b^2 + (t-t_1)^2} 
\end{equation}
with highest value at $t = t_1$ yielding the fitting parameters $(\Delta{S_{max}})_0$ = 1.49 JKg$^{-1}$K$^{-1}$, $a$ = 805.15 JKg$^{-1}$K$^{-1}$nm$^2$, $b$ = 20.3 nm, $t_1$ = 66.3 nm for LCMO/LAO (Fig. 6(c)) and $(\Delta{S_{max}})_0$ = 4.14 JKg$^{-1}$K$^{-1}$, $a$ = 3124.23 JKg$^{-1}$K$^{-1}$nm$^2$, $b$ = 18.4 nm and $t_1$ = 66.8 nm for LCMO/STO (Fig. 6(d)). From the Lorentz function fitting we observed that the value of $t_1$ ($\sim$ 66 nm ± 5nm) is almost same for both LCMO/LAO and LCMO/STO. Certainly more data points could give precise value of $t_1$ by reducing the deviation of the parameters.
\begin{figure}
\centering
\includegraphics[width=8.8cm]{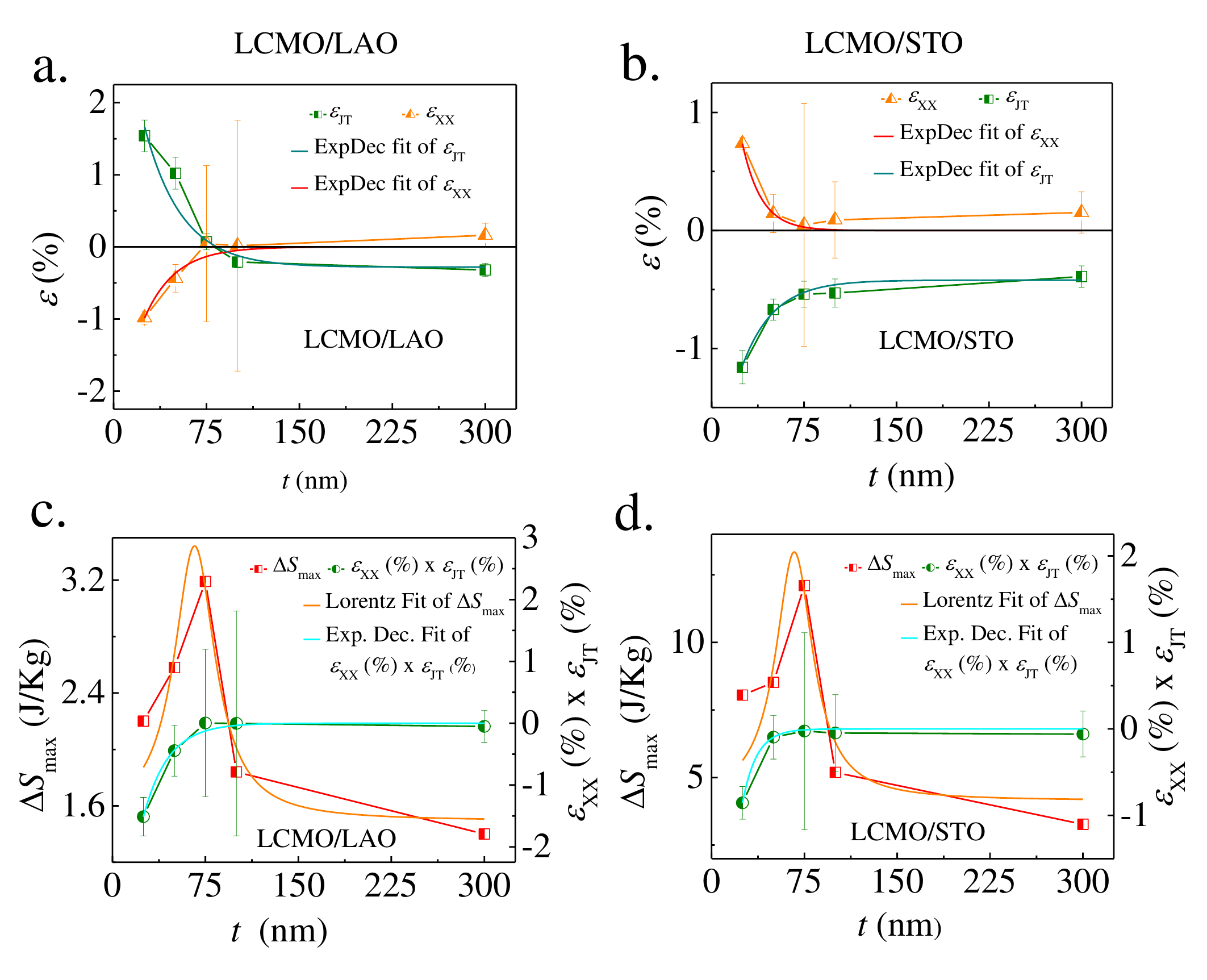}
\caption{In-plane biaxial strain ($\epsilon_{XX}$) relaxation and decaying of Jahn-Teller distortion ($\epsilon_{JT}$) of (a) LCMO/LAO and (b) LCMO/STO. The vertical error bars are standard deviation. (c) and (d) $\epsilon_{XX}$ $\times$ $\epsilon_{JT}$ and $\Delta{S_{max}}$ vs $t$ plot of LCMO/LAO and LCMO/STO, respectively.}
\end{figure}

\begin{table}
\caption{Comparison of $\Delta{S_{max}}$ (in JKg$^{-1}$K$^{-1}$) and RCPs (in JKg$^{-1}$) of different MC materials}
\begin{ruledtabular}
\begin{tabular}{ccccc}
\textrm{Compound}&
\textrm{$\Delta{S_{max}}$}&
\textrm{RCP-1}&
\textrm{RCP-2}&
\textrm{Ref}\\
\colrule
Pr$_5$Ni$_{1.9}$Si$_3$ & 8.15\footnote{Under the field change of 5 T\label{a}} & 165 & - & \cite{pecharsky2003preparation} \\
PrNi & 6.15\footref{a} & 56 & - & \cite{pecharsky2003preparation} \\
Er$_3$Ni$_2$ & 19.5\footref{a} & 507 & - & \cite{dong2011magnetic} \\
TmGa & 34.2\footref{a} & 485 & 364 & \cite{mo2013low} \\
GdPd$_2$Si & 6.0\footref{a} & 329 & - & \cite{rawat2001magnetocaloric} \\
ErFeSi & 23.1\footref{a} & 460 & 365 & \cite{zhang2013large} \\
LCMO/STO & 12.1\footnote{Under the field change of 6 T\label{b}} & 361 & 257 & This Work \\
LCMO/LAO & 3.2\footref{b} & 339 & 258 & This Work \\
\end{tabular}
\end{ruledtabular}
\end{table}\textbf{}

\begin{figure}
\centering
\includegraphics[width=8.8cm]{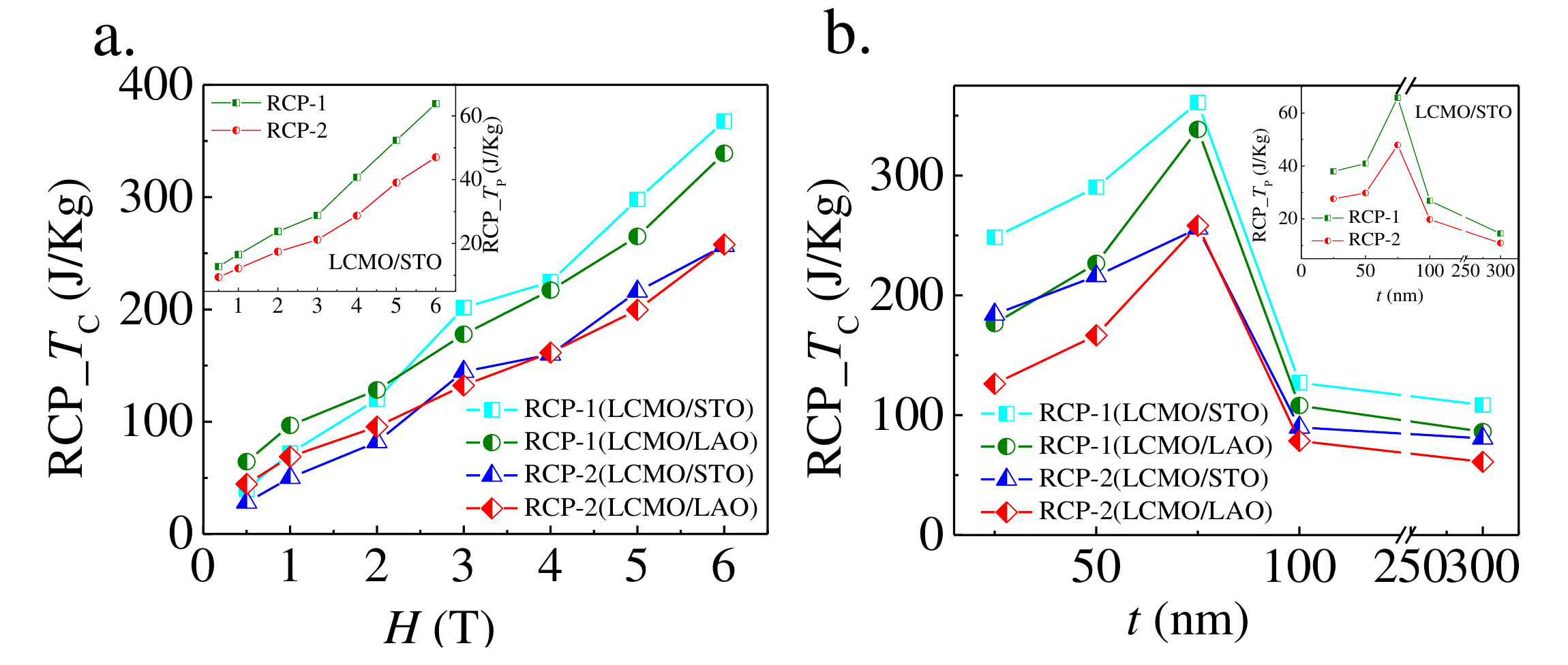}
\caption{RCP for LCMO/STO and LCMO/LAO calculated around $T_C$ as a function of (a) $t$ (b) $H$. Inset figures show the RCPs of LCMO/STO calculated around $T_P$ as a function of $t$ and $H$, respectively.}
\end{figure}

Generally, the first order transition materials (FOTM) undergo larger magnetic entropy change ($\Delta{S_M}$) with the smaller working temperature span (WT span) than the second order transition materials (SOTM) and exhibit temperature dependent magnetic hysteresis which diminishes the energy efficiency of the magnetocaloric materials\cite{chaudhary2015high}. So, confusion may arise to figure out the suitability in cooling application between FOTM with larger $\Delta{S_M}$ but smaller WT span and SOTM with smaller $\Delta{S_M}$ but larger WT span. To overcome this, instead of focusing only on WT span or only on $\Delta{S_M}$, the net heat extracted in a cooling cycle known as RCP is calculated\cite{chaudhary2015high}. Two methods can be used to calculate RCP\cite{provenzano2004reduction}, 1st method - Wood and Potter method\cite{wood1985general} by which RCP is defined by 
\begin{equation}
RCP-1 = \Delta{S_M}\Delta{T}
\end{equation}
where $\Delta{S_M}$ is the magnetic entropy change at the hot and cold ends of the cycle (defined equal) and $\Delta{T}$ = $T_{hot}–T_{cold}$ and the 2nd method - the area under the $\Delta{S_M}$ vs $T$ curves taking the temperatures of the cold and hot sinks as lower and upper limit of integration, respectively\cite{gschneidner1999recent}.In both methods, we take $\delta{T}_{FWHM}$ as the working temperature range following the protocol as $\Delta{S_M}$ almost vanishes beyond this temperature range\cite{zverev2010maximum,caballero2011optimization}. So,
\begin{equation}
     RCP-2 = \int_{T_{cold}}^{T_{hot}} \Delta{S_M}(T,H_{max}) \,dT [50]
\end{equation}
where $T_{hot}–T_{cold}$= $\delta{T}_{FWHM}$ and $H_{max}$ is the maximum value of the applied field.
It shows a large RCP-1 value, i.e., ~ 361 Jkg$^{-1}$ for LCMO/STO at $H$ = 6 T. The estimated values of RCP-2 and RCP-1 near $T_C$ as a function of $t$ and $H$ are plotted in Figs. 7(a) and 7(b), respectively. Insets of Fig. 7 show the RCPs around $T_P$ for LCMO/STO.

It can be observed that for both the temperature regimes, the value of RCP linearly increases with increasing $H$ and has the largest value at $t$ $\sim$ 75 nm. In practical cooling applications, the material in the same refrigeration cycle with higher RCP is preferred as it would confirm the transport of a greater amount of heat in an ideal refrigeration cycle. To make the applicability of our results, i.e., to make the LCMO thin film as a magnetic refrigerant, $\mid$$\Delta{S_{max}}$$\mid$ and the values of RCPs, i.e., RCP-1 and RCP-2 are determined in this present study and they are compared in Table-I with several other magnetic refrigerants reported earlier in literatures. It is clearly evident from the Table-I that the obtained peak values of $\mid$$\Delta{S_{max}}$$\mid$ and the RCPs at a critical $t$ of 75 nm in our work are also comparable with other listed magnetic refrigerants.\\

\section{Conclusion}
We reported an anomalous strain effect in LCMO/STO and LCMO/LAO showing that MCE as well as transition temperature for both the tensile and compressive strain at first increases with $t$ due to in-plane strain relaxation showing maximum values at $t$ $\sim$ 75 nm and then, it decreases. To explain the anomaly of increasing MCE for both the compressive and tensile strain, two well-known effects have been considered simultaneously- D-E interaction between Mn$^{3+}$ and Mn$^{4+}$ ions and $3d_{x^2-y^2}$ orbital stabilization. We have also shown the in-plane strain ($\epsilon_{XX}$) relaxation and decaying of JT distortion ($\epsilon_{JT}$). $\epsilon_{XX}$ , $\epsilon_{JT}$ and the product ($\epsilon_{XX} \times \epsilon_{JT}$) for LCMO/STO decay more rapidly than those of LCMO/LAO, respectively. These decay are exponential unlike $t$ dependence of $\Delta{S_{max}}$ which is the Lorentz function with the maximum value at $t$ $\sim$ 66±5nm for both LCMO/LAO and LCMO/STO. Here, $t$ can be optimized for larger MCE using this Lorentz function. Moreover, the variation in transition temperatures with respect to $t$ and substrates makes the LCMO thin films flexible for using at different operating temperatures.
\section*{Acknowledgement}
WA acknowledges DST-INSPIRE. MB acknowledges institute funding from IISER Thiruvananthapuram.

\bibliographystyle{unsrt}
\bibliography{bibliography.bib}
\end{document}